\documentclass[review]{elsarticle} 

\usepackage{makecell} 
\usepackage{todonotes} 

\usepackage{longtable}
\usepackage{amsmath} 

\usepackage{graphicx}
\usepackage{subcaption}

\newcommand{\R}{\ensuremath{\mathbb{R}}}

\DeclareMathOperator*{\argmin}{arg\,min}

\usepackage{graphics} 

\usepackage{soul} 
\usepackage{ulem}

\usepackage{url} 
\usepackage{amssymb}
\usepackage{mathtools} 

\title{Enabling Personalized Decision Support with Patient-Generated Data and Attributable Components}

\author[1]{Elliot G Mitchell\corref{cor1}} \ead{egm2143@cumc.columbia.edu}
\author[2]{Esteban G Tabak} \ead{tabak@cims.nyu.edu}
\author[3]{Matthew E Levine} \ead{mlevine@caltech.edu}
\author[1]{Lena Mamykina} \ead{om2196@cumc.columbia.edu}
\author[1,4]{David J Albers} \ead{david.albers@ucdenver.edu}

\cortext[cor1]{Corresponding author. \textit{Email:} egm2143@cumc.columbia.edu}

\address[1]{Department of Biomedical Informatics, Columbia University, New York, NY, USA}
\address[2]{Courant Institute of Mathematical Sciences, New York, NY, USA}
\address[3]{California Institute of Technology, Pasadena, CA, USA}
\address[4]{Department of Pediatrics, Division of Informatics, University of Colorado, Aurora, CO, USA}

\begin{document}

  \begin{abstract}

    Decision-making related to health is complex. Machine learning (ML) and patient generated data can identify patterns and insights at the individual level, where human cognition falls short, but not all ML-generated information is of equal utility for making health-related decisions. We develop and apply attributable components analysis (ACA), a method inspired by optimal transport theory, to type 2 diabetes self-monitoring data to identify patterns of association between nutrition and blood glucose control. In comparison with linear regression, we found that ACA offers a number of characteristics that make it promising for use in decision support applications. For example, ACA was able to identify non-linear relationships, was more robust to outliers, and offered broader and more expressive uncertainty estimates. In addition, our results highlight a tradeoff between model accuracy and interpretability, and we discuss implications for ML-driven decision support systems.

    \centering
    \includegraphics[width=4.5in]{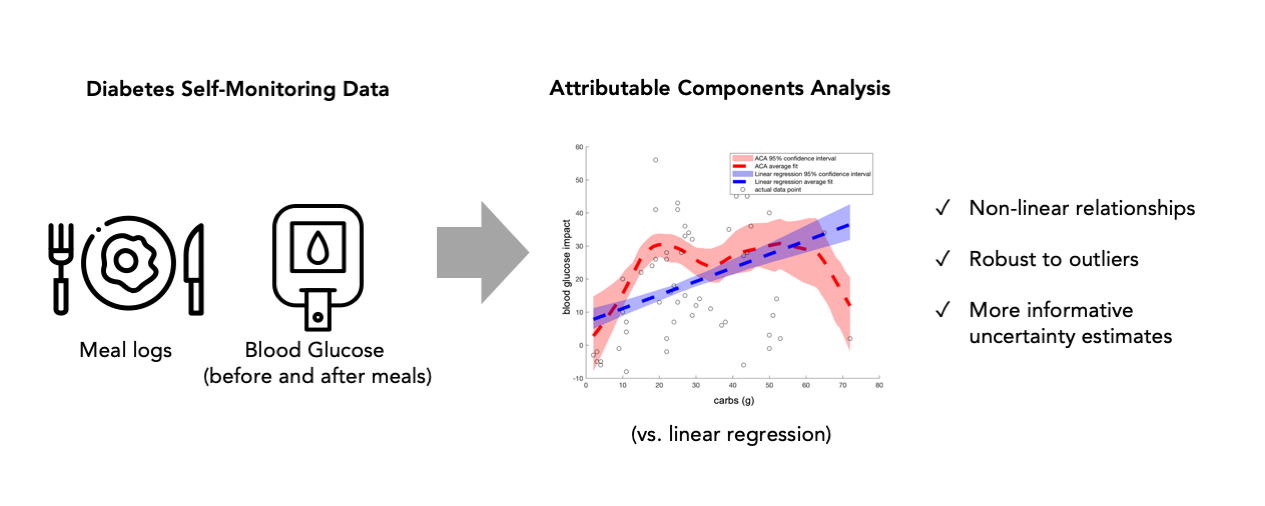}
    \textbf{Graphical Abstract}

  \end{abstract}

\maketitle

\section{Introduction}

In complex domains like health, it can be difficult to anticipate the consequences of daily choices on short- and long-term health status. Collecting and analyzing data about behaviors and indicators of health can elucidate patterns of association between a behavior and a range of outcomes. Thanks to wearable sensors and mobile health applications, patient-generated health data can be collected more easily than ever, but questions remain about how to incorporate these data into health decisions \citep{Genes2018FromData,2017MobileHealth,Fiordelli2013MappingEvolution}.

One area where patient-generated data holds promise to inform decision-making is type 2 diabetes self-management. In type 2 diabetes, a key goal of self-management is keeping blood glucose (BG) within target ranges. Daily behaviors like diet have a direct relationship with BG levels. Importantly, different individuals have different glycemic responses to different foods \citep{Zeevi2015}, emphasizing a need for personalization \citep{AmericanDiabetesAssociation2018}. Estimating the impact of a meal on BG is difficult, even for experts \citep{Mamykina2016Data-drivenTechnologies}. Machine learning (ML) may be better suited to identify consistent patterns than human reasoning  \citep{Albers2017PersonalizedAssimilation}.

Using patient-generated data for personalized analysis in the context of nutrition and BG, however, poses challenges. BG measurements and meals need to be actively tracked by users, which requires effort. Fingers need to be pricked to record BG, and meal details need to be entered. Because of the burden of entry, these data points are incomplete and non-randomly missing \citep{Cordeiro2015BarriersNudges}. With nutrition logging in particular, there is a tradeoff between the time and effort of logging and the detail and accuracy of the nutrition information logged \citep{Cordeiro2015,Andrew2013SimplifyingDiaries}. Gold standard nutrition evaluations require analysis in a specialized lab, which is unavailable for patient-generated meal logs.
In addition, glucometers can be miscalibrated, and users can mistype entries leading to both systematic bias and random errors. Glucose dynamics themselves are non-linear, oscillatory, noisy, and depend on individual characteristics \citep{Ismail-Beigi2012GlycemicMellitus,Albers2017PersonalizedAssimilation}.
Similar to the data quality concerns of electronic health records, the incompleteness, inaccuracy, complexity, and bias of patient-generated data create challenges for accurately representing a patient's state \citep{Hripcsak2013Next-generationRecords.,Weiskopf2013MethodsResearch}.
Still, prior work has demonstrated that accurate inference can be possible with similar data sets \citep{Albers2017PersonalizedAssimilation, Albers2018MechanisticPhenotype}

In addition to the challenges of the data, though, designing analysis for decision support tools brings its own substantial challenges. Algorithms need to be able to run as a part of an automated system, identifying complex relationships while being robust to outliers. In addition, it's important for the output to be interpretable.
By interpretable, we mean that the relationships identified in the output of the model can be translated into useful and actionable support for decision-making.
Notably, this definition diverges from ``interpretable'' as similar to ``explainable'' ML, which seeks to disentangle predictions from deep learning and other black box models {\citep{gunning2017explainable, holzinger2017we}}.
Interpretability is important because even the most accurate ML machinery is not useful if it cannot affect decision-making or be transformed into an understandable action. Quantifying uncertainty is an important part of interpretability, so that model output can be weighed appropriately in the decision-making process \citep{Cabitza2019AMedicine, Cabitza2017UnintendedMedicine}.

There is a need for methods that address these challenges. Optimal transport is a theory that offers tools to estimate and compare probability distributions \citep{Peyre2018ComputationalTransport,Villani2009OptimalNew}. In its original formulation, optimal transport sought to optimize the transportation of goods and resources, but has since been applied to many problems like computer vision and machine learning \citep{Peyre2018ComputationalTransport}.
Optimal transport is particularly useful for data where values are highly individualized, as in medicine \citep{Albers2014DynamicalPopulations}.
Blood pressure, for instance, may be related to many factors like age, exercise, diet, sex, prescribed drugs, and the device used to take the measurement.
Here we adapt a optimal transport-based method invented by  \citet{tabak-trigila2018AC} termed attributable components analysis (ACA).
This method was created to explain variability in a quantity of interest based on a set of related or potentially confounding covariates, or ``attributable components''. Each component represents a contribution to the observed variability while simultaneously filtering out irrelevant effects to focus on a particular relationship.

Here, we apply an adapted version of the ACA method to type 2 diabetes self-monitoring data, using ACA to estimate the mean glycemic impact of a meal---the difference between pre-meal and post-meal measurements---based on the meal's macronutrient composition. By estimating how each attributable component, in this case each macronutrient, contributes to the variability in BG after a meal, ACA can identify patterns of association between each macronutrient and expected BG impact. To better understand and convey how ACA performs for this task, we compare its output to linear regression. We then discuss how these estimates can be used as input to decision support systems, for example finding personalized ranges of macronutrient values where BG impact is expected to be higher or lower, to inform clinical care or create personalized nutrition plans.

\section{Materials and Methods}

\subsection{Data Set}
\label{sec:dataset}

The data used in this research originates from prior user studies of a smartphone application for diabetes self-monitoring. In the application, participants logged meals and blood glucose (BG) readings. To log a meal, users captured a photograph of the meal, assigned a category of the meal (breakfast, lunch, dinner, or a snack) and entered a free-text description of the meal contents. Users entered pre-meal BG readings when logging the meal. Two hours after each meal, users received a notification to record and enter their post-meal BG reading. Later, each meal was evaluated by a registered dietitian (RD) who performed a nutrient assessment of the meal using a standard protocol and the USDA food composition database \citep{USDAUSDADatabase,Ahuja2013USDAPractice}. The RD recorded the carbohydrates, fat, protein, and fiber, in grams, as well as the total calories of the meal.

Data came from 40 users who used the smartphone application for 4 to 12 weeks in a separate IRB approved study. Each participant consented for their data to be re-used in future research. In this analysis, we included all participants with 30 or more total meals logged, and considered only the meals with both pre- and post- meal BG readings, for a total of 16 users.

\subsection{Descriptive Statistics}

\begin{figure}
  \centering
  \includegraphics[width=4in]{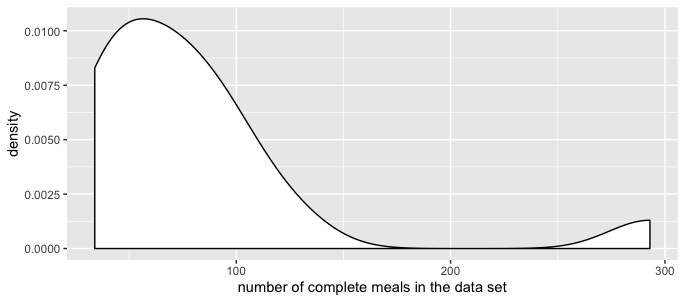}
  \caption{Kernel density estimate of the number of users with n-many meals in the data set. The mass of the distribution sits near the median of 67 meals loggged, with a long tail of users logging considerably more meals.}
  \label{fig:complete_meals_dist}
\end{figure}

\begin{figure}
  \centering
  \includegraphics[width=3.25in]{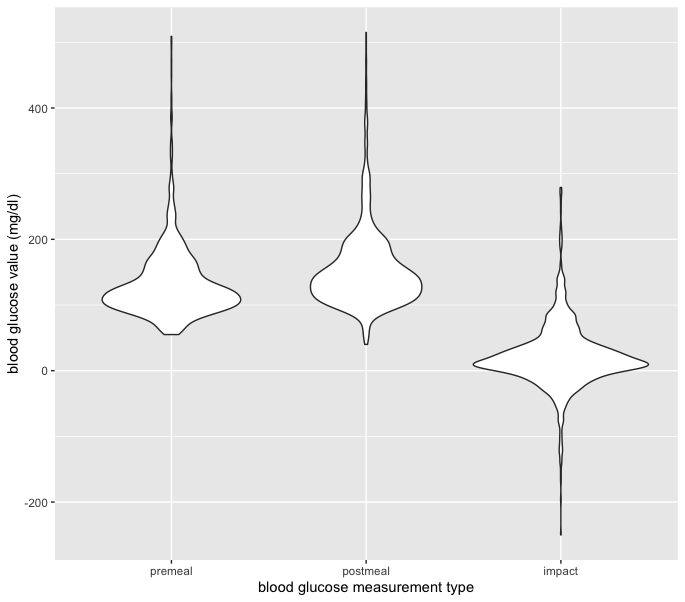}
  \caption{Violin plots showing the distribution of blood glucose readings across all users. Users varied considerably in their blood glucose levels before and after meals.}
  \label{fig:bg_dist}
\end{figure}

The 16 users with type 2 diabetes collected a median of 67 meals over 4 to 12 weeks. As seen in Figure \ref{fig:complete_meals_dist}, most users logged close to the median number of meals, with a few users logging considerably more. As shown in Figure \ref{fig:bg_dist}, users varied substantially in their BG levels before and after meals.

\begin{figure}[htbp]
  \centering
  \includegraphics[width=4in]{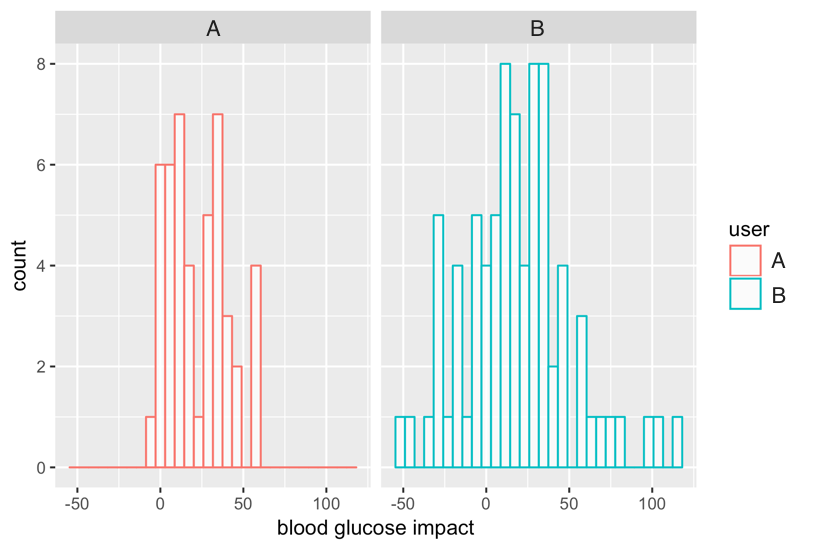}
  \caption{A histogram of BG impacts for users A and B. User A had less variability in BG impacts compared to user B.}
  \label{fig:bg_impacts}
\end{figure}

\begin{figure}[htbp]
  \centering
  \includegraphics[width=4in]{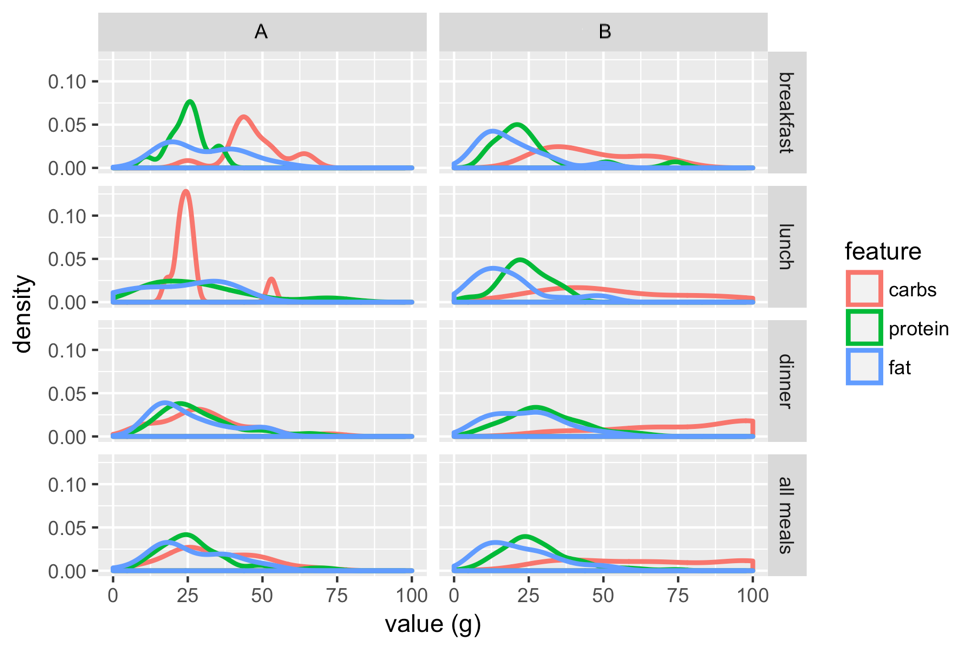}
  \caption{Kernel density estimate plots of macronutrient consumption for users A and B.
  There is variability in macro consumption between and within each user. Note that nutrition evaluations only allowed up to 100 grams of each macronutrient, and user B regularly ate 100 grams or more of carbohydrates at dinner.}
  \label{fig:macro_density}
\end{figure}

Two users, ``A'' and ``B'', were chosen for a detailed inspection of model performance because they were representative of the overall data set, but differed from each other in BG control and macronutrient consumption patterns. Users A and B logged a total of 58 and 88 meals over 4 and 12 weeks, respectively. See Table \ref{tab:meal_counts} for a detailed breakdown by meal type. As seen in Figure \ref{fig:bg_impacts} user A had less variability in BG impacts compared to B. Figure \ref{fig:macro_density} shows kernel density estimates of the macronutrient features for both users. Shown side by side, these densities show variability between and within each user. For example, user A ate 25 grams of carbohydrates at lunch most of the time, while user B had much more variability in their lunchtime carbohydrate intake. An important artifact and limitation is that nutrition evaluations only accommodated up to 100 grams of each macronutrient to be entered, yet user B regularly ate 100 grams or more of carbohydrates at dinner.

\begin{table}[htbp]
    \centering
    \caption{Count of meals of each meal type for users A and B.}
    \begin{tabular}{|l|l|l|}
        \hline
        User ID & Meal Type & Count \\ \hline
        A      & Breakfast & 13    \\
                & Lunch     & 10    \\
                & Dinner    & 23    \\
                & Other     & 12    \\
                & Overall   & 58    \\ \hline
        B    & Breakfast & 16    \\
                & Lunch     & 19    \\
                & Dinner    & 44    \\
                & Other     & 9     \\
                & Overall   & 88    \\ \hline
    \end{tabular}
    \label{tab:meal_counts}
\end{table}

\subsection{Feature Selection}
We experimented with different representations of features to predict BG impact. We began with the three main macronutrients---carbohydrates, fat, and protein---represented as their weight in grams, or their proportion of each meal's calories. ACA performed slightly better when representing macronutrients as proportions than as grams, but we opted to use grams because we thought this would be more useful for decision support. In an effort to make decisions more straightforward, nutrition education in diabetes emphasizes the importance of macronutrients, and usually focuses on amounts of foods with units like grams, not their contribution to calories \citep{wheeler2014choose}. While some materials like the USDA's MyFoodPlate are based on the proportion of the plate filled with different foods, the proportion of calories is very different than the volume a food takes up on a plate. (Consider 1 stick of butter vs. 4 cups of raw spinach.) And finally, representing macronutrients as proportions means that the values sum to one, which introduces strong multicollinearity that creates challenges for inference with linear regression.

In addition, we also included fiber and pre-meal BG as features. We included fiber because increasing fiber is a common recommendation for individuals with diabetes \citep{Anderson2004CarbohydrateEvidence}. We included pre-meal BG because of its relationship with post-meal BG. Glucose dynamics at their simplest consist of a glycemic response to nutrition. Because of this, to infer glycemic response to nutrition---to solve the equations uniquely---we need the initial state (pre-meal glucose), the kick (nutrition consumption), and the response (post-meal glucose).

A particular challenge of type 2 diabetes self-monitoring data is representing the impact of a particular meal on BG, or the glycemic impact. An optimal sampling rate for BG is on the order of minutes, not hours \citep{Breton2008OptimumMonitors.,Gough2003FrequencyDynamics}.
A single reading two hours after the meal is the clinical standard for postprandial measurement \citep{Aschner2017NewCare} but is not well suited to capture the fluctuations in BG after a meal. Even with appropriately sampled continuous glucose monitoring (CGM) data, it's not clear which features are most important to diabetes-related complications; the highest peak in blood glucose, the integral of the glycemic curve from the mean to some time after the meal, the average value over time, or the speed of oscillations following a meal are different ways of representing BG impact, with different potential physiologic implications. While more frequent or continuous measurement would be preferred from a data standpoint, checking BG 6-10 times per day is recommended for those on insulin therapy, and there is no recommendation for those not on insulin \citep{AmericanDiabetesAssociation2018}.  Here, we follow the standard practice for postprandial BG measurement, and take the difference of post-meal BG minus pre-meal BG to represent the glycemic impact of a meal.


\subsection{Attributable Components}
\label{sec:MO}

Attributable component analysis \citep[ACA;][]{tabak-trigila2018AC} is a methodology for explaining the potentially nonlinear variability in a quantity of interest, $x$, in terms of covariates $z=\left(z_1, \ldots, z_L\right)$.
The method is highly motivated by theory and ideas from optimal transport \citep{Villani2009OptimalNew,Santambrogio2010IntroductionTheory}.
In our application, $x$ represents the glycemic impact, and $z$ for the macronutrient content of a meal. The covariates can be categorical (such as ``meal'', with values in $[\hbox{``breakfast'', ``lunch'', ``dinner''}]$), real (such as ``total amount of carbohydrates'') or, in fact, of nearly any type. The output of attributable component analysis is $\bar{x}(z)$, the conditional expectation of $x$ with respect to covariates $z$; this conditional mean is provided as a sum of components, which can be thought of as modes of variability. Each component is represented by the product of one-dimensional functions of each covariate $z_l$.

A more detailed explanation of ACA is provided in  \ref{appendix_a}, but a summary is provided here.

Given a set of $m$ observations of the variable of interest $x$ and $L$ covariates, $\bigl\{ \{z^{(i)}_l\}_{l=1}^L,x^{(i)} \bigr\}_{i=1}^m$, the ACA algorithm seeks to estimate the conditional mean $\bar{x}(z)$ with the following equation:

\begin{equation}
  \bar{x}(z_1, \ldots, z_L) = \sum_{k=1}^d \prod_{l \in L} \sum_j \alpha(l)^j(z_l) V(l)_j^k,
  \label{eq:aca}
\end{equation}

 each $k$ is a component of the variability in $x$, the $V$'s are essentially basis functions that represent the variability, and can be represented by many classes of functions, e.g., as the sum of the product of sinusoidal functions in the case of Fourier decomposition (cf. Appendix \citep[ACA;][]{tabak-trigila2018AC}), and $\alpha(l)_i^j = 1$ when $z_l^i = j$ and $\alpha(l)_i^j = 0$ otherwise.

The complete estimate of $\bar{x}$ based on all $L$ features is useful, but being a probability distribution, is difficult to translate into useful recommendations because of the complexity dimensionality.  To address this problem, we instead use the marginal dependence that translates $\bar{x}$ from an $L-$dimensional function into a one dimensional function.



\subsubsection{Interpretability through marginalization}
\label{sec:marginals}

We make the ACA output more interpretable for decision-making by ``marginalizing'' the ACA output function. To understand what this means, why this is necessary, and how this works, begin with the ACA estimated conditional mean that adopts the form in Equation \ref{eq:aca}
where the $V(l)_j^k$ are found by the algorithm, and the $\alpha(l)^j(z_l)$ are known via interpolation on grids or prototypal analysis. Even though this estimation allows us to make predictions for new values of $z$, its complexity makes it difficult to interpret. For example, if we limit the covariates to only binary forms, e.g., increases or decreases, then there are $2^L$ combinations of actions a person must interpret and choose among; this is too complex. Because the point of this intervention is to help people understand glycemic impacts of nutrition to make balanced choices that are sustainable behaviorally, we must translate ACA output into a simpler form, one where the impact of a single covariate is considered at a time, leading to only $L$ different options.  We can do this by asking simpler questions, such as: averaging over all other covariates, how does $x$ depend on a specific $z_l$ or small set thereof.  Such questions ask us to marginalize the full estimated conditional mean and the separated form of the estimation makes it straightforward to perform this task.
In order to find the marginal dependence of $x$ on a group of covariates $H$ denoted by $\{z_{h_t}\}_{t = 1}^{s}$, with $h_t \in H$ and $s = |H|$, one has

$$  \bar{x}(z_{h_1}, \ldots, z_{h_s}) = \hfill $$
\begin{equation}
    \sum_{k=1}^d \left[\frac{1}{m}\sum_{i=1}^m \prod_{h \not{\in} H} \sum_j \alpha(h)_i^j V(h)_j^k\right]\prod_{h \in H} \sum_j \alpha(h)^j(z_{h}) V(h)_j^k .
\label{equation:margin}
\end{equation}

In this case, $ \bar{x}(z_{h_1}, \ldots, z_{h_s})$ represents a function that captures the impact of a particular subset of features on $x$. For a single covariate of interest $h$, $\bar{x}(z_h)$ is a one-dimensional function that captures the impact that one covariate, for example fat, has on glycemic impact.  In Figs. \ref{fig:nonlinearity}, \ref{fig:outliers}, and \ref{fig:uncertainty} where we compare the ACA to linear regression, the one-dimensional ACA output shown is $\bar{x}(z_h)$ as opposed to the full ACA model $\bar{x}(z_1, \ldots, z_L)$.

\subsection{Other regression methods and ACA}

There are other methods that can be used for similar tasks. ACA is a non-parametric density estimation method, and its task of explaining variability based on a set of covariates is similar to regression with clustering or principal components analysis (PCA). Importantly, ACA's output is more interpretable than these alternatives. If the goal is to identify patterns between an individual's nutrition and their glycemic control or to make recommendations to change diet, then it's important that the output can be translated for human understanding. With ACA, each attributable component is a covariate, meaning the relationships identified are in the same dimensions as the input data. PCA finds the uncorrelated components that explain the most variability in the dependent variable {\citep{Jolliffe2016PrincipalDevelopments.}}, but what exactly each component means could be difficult to explain in a clinical situation. Similarly, clusters can be difficult to convey to clinicians without extensive training, and require interpretation {\citep{Feller2018AData}}. It's important that the model output aligns with cognitive models {\citep{Pazzani2001AcceptanceExperts}}; a complex, black box method with strong performance metrics is only useful if it can be translated into something clinically meaningful.

As ACA is operationalized here, its output is also similar to other regression methods like least-squares or support vector machine (SVM) regression.
However, it's notable that the method by which ACA estimates this regression is by approximating a joint distribution and marginalizing over the features, which is different than how other methods fit the data.

In choosing a comparison method, we aim to identify and highlight qualitative and quantitative differences between ACA and another regression approach. We do not aim to argue for the hypothesis that ACA is the {\textit{best}} method for this data and task, and an intrinsic evaluation of ACA has been reported elsewhere and is outside the scope of this work. As a baseline, therefore, we compare ACA against multiple linear regression {\citep{Mendenhall1997AAnalysis}}.
While there are many potential choices for a regression comparator, including various non-linear variants, linear regression is a highly used model and is a reasonable choice for our data because its limited complexity means it has the potential to perform well on small, n-of-1 data sets in our experiments.

\subsection{Comparator: Linear Regression}

As a comparison method, we fit the data with multiple linear regression
\begin{equation}
  \bar{x}(z_1,...,z_L) = \beta_0 + \beta_1 z_1 + \beta_2 z_2 +...+ \beta_L z_L
\end{equation}
where $x$ is the quantity of interest and $z_1,...,z_L$ are covariates and $\beta_0$ is the intercept term. More compactly
\begin{equation}
  \bar{x}(z_1,...,z_L) = \beta_0 + \sum_{l=1}^L \beta_l z_l
\end{equation}
We then find the best fit using the ordinary least squares method \citep{Mendenhall1997AAnalysis}.

As with ACA, to improve the interpretability of the output, we fit the model with all covariates, $z$, but marginalize to consider a specific $z_l$ (or small subset) by averaging over the other covariates. To compute the marginal dependence of $x$ on a group of covariates $H$ denoted by $\{z_{h_t}\}_{t = 1}^{s}$, with $h_t \in H$ and $s = |H|$, one has

\begin{equation}
  \bar{x}(z_{h_1}, \ldots, z_{h_s})
  = \beta_0 +
    \sum_{h \not{\in} H} \beta_h \left[\frac{1}{m} \sum_{i=1}^m z_{h}^{(i)} \right]  +
    \sum_{h \in H} \beta_h z_h
\label{equation:lm_marginal}
\end{equation}

\subsection{Translating Inference-Based Analysis -- ACA and Linear Regression -- to Decision Support} \label{regression2cds}

The outcome of the marginalization calculation in Eq. {\ref{equation:margin}} and the linear regression in Eq. {\ref{equation:lm_marginal}} is a one-dimensional graph, e.g., Figure {\ref{fig:nonlinearity}}, where the macronutrient is given on the x-axis as the independent variable or covariate and the y-axis is the glycemic impact. This plot is not, alone, useful for making decisions for most patients, clinicians, or machines.

{The aim is to support patients in making decisions about how to modify their diet to improve their BG levels, for example increasing or decreasing the amount of a macronutrient like protein in their diet. To be useful to these ends, the raw information output by ACA needs to be translated and interpreted with contextual information and clinical knowledge. For example, this clinical knowledge might include an understanding of what constitutes a ``good'' or ``bad'' BG impact, or what gradations of good/bad BG impact are and at what resolution.}

{One simple approach would be to determine a clinically significant threshold for BG impact that is too high, and aim to keep individuals below that threshold. Then one draws a horizontal line to identify ranges of each macronutrient where mean BG impact is expected to be above or below the threshold. These ranges could be useful for patients, educators, or providers in setting a personalized nutritional plan {\citep{AmericanDiabetesAssociation2018}}, or as input to another system that recommends recipes or meal plans with nutritional constraints, with the aim of reducing the number of BG impact excursions above the determined threshold.}

{Building on a simple approach, the intricacies of the relationships identified by ACA could also be analyzed more flexibly, for example being interpreted by a rule-based expert system to find target levels of macronutrient consumption that would minimize BG impact for an individual, within a set of constraints.}

{While we do not go all the way to translating ACA output into a decision support system in this paper, we introduce these concepts to provide context for the evaluation and discussion that follow. There are many possibilities for how the output of a method like ACA could be interpreted within a decision support system, and we aim to explore the properties of the algorithm for use in such a task.}

Notably, there are other approaches for personalized decision support in health. For example, a growing body of research in health-focused recommender systems aims to help individuals choose healthy meals by suggesting meals that are likely to be desirable, but fall within a set of health constraints \citep{Yang2016YummeSystem,Elsweiler2016EngenderingSystems}. While these systems also rely on data and ML for personalization, recommender system algorithms are intended to learn user preferences, not health constraints.
In nutrition, personal preference and health can often be in tension, and in these systems, the constraints for what makes a healthy meal are {\textit{not}} personalized \citep{Yang2016YummeSystem,Elsweiler2016EngenderingSystems}.
ACA would be complementary to these approaches, and could be applied alongside a health-focused recommender system to learn personalized macronutrient constraints within a recommender system that aims to suggest desirable meals.

\subsection{Uncertainty Estimates} \label{uncertainty}

We used several bootstrapping algorithms to estimate uncertainty of the regressions. Specifically, we used bootstrap to estimate distributions of regression coefficients, allowing us to quantify the variability of the estimate. Given this distribution we can calculate quantities that characterize the uncertainty; here we focus on confidence intervals over the range of input values.  Often, bootstrapping is accomplished by drawing multiple samples with replacement from the data set and computing the estimate for that resampled data \citep{Davison1997BootstrapApplication}. Empirical confidence intervals can be calculated from the distribution of estimates. In addition, ACA is stochastic, with a random initial state, so we can estimate the variability through repeated calculations with the same subset but different starting states. We experimented with both methods for bootstrapping ACA, and the results were nearly identical. We opted for the typical approach of bootstrapping via multiple subsamples so that we could apply the same bootstrapping procedure for both methods, because linear regression is not stochastic.

A second question is the size of the bootstrap samples. A common approach is for each bootstrap sample to have the same number of data points as the original data set. Because data sets for some of the users were quite small, there were advantages to using larger bootstrap samples. For example, bootstrap samples may have very few unique data points. This negatively impacts the performance of the model, and poses challenges for aggregating variance estimates across the complete range of feature values. Larger bootstrap samples can improve model performance, and help ensure that estimates cover the full range of independent variable values; of course bootstrap ensembles cannot represent the tails of distributions that are not observed in the data, and can underestimate variance. We experimented with the original size of the dataset, 100, and 500 data points, and found that a bootstrap sample size of 500 performed well for both ACA and regression.

A third question is how many bootstrap iterations to run. 100 iterations has been suggested as a minimum for variance estimations, but it depends on the situation \citep{Davison1997BootstrapApplication}. We inspected the change in variance across all iterations after each subsequent bootstrap iteration to look for convergence. We experimented with up to 200 iterations and found that 100 iterations were sufficient for variance to converge.

All analysis was performed in MATLAB 2016b (9.1). Additional plots and descriptive statistics were produced in R v3.3.2 with tidyverse v1.1.1.

\subsection{Experimental Design}

We estimated ACA and linear regression on the data sets for each user, as well as data subsets by meal type (breakfast, lunch, and dinner). To estimate confidence intervals, we performed a bootstrap with 100 iterations, based on the procedure described in section \ref{uncertainty}. Each bootstrap sample had 500 data points, and the same samples were used to fit ACA and linear regression. 95\% confidence intervals were determined empirically from the aggregated bootstrap output.

We then produced a series of plots for each user and closely inspected the plots for the two users introduced in section \ref{sec:dataset}. Each plot included an individual covariate $(z_l)$ on the horizontal axis, with BG impact $x$ on the vertical axis, the actual data points, and average fit of ACA and linear regression with confidence intervals. With each of the 5 features for the overall data sets and the 3 meal-type subsets across two users, there were a total of 40 plots. See Figure \ref{fig:nonlinearity} in the Results for an example.

\subsection{Evaluation}

To compare the performance of the two models we calculated the root mean squared error (RMSE) of the data fit for both ACA and linear regression.

RMSE for the overall model:
\[\sqrt{\frac{1}{m}  \sum_{i=1}^m |\bar{x}(z^{(i)}_1, ..., z^{(i)}_L) - x^{(i)}|^2}\]

RMSE for the marginals:
\[\sqrt{\frac{1}{m} \frac{1}{L} \sum_{l=1}^L \sum_{i=1}^m |\bar{x}(z^{(i)}_l) - x^{(i)}|^2}\]

In addition, we qualitatively inspected the plots for evidence of non-linear relationships, and examined the situations where the two models agreed and disagreed. To quantify non-linear relationships, we heuristically evaluated the plots to tally the number of data sets where the average fit line of ACA had more than a 10-degree bend.

To quantify differences in the uncertainty calculations between the two methods, and to assess the coherence and usefulness of the confidence intervals, we calculated the percentage of data points falling within the confidence interval across all data sets.

\section{Results}

\subsection{Evaluation}

As shown in Table \ref{tab:rmse_all}, the RMSE for the full ACA model was significantly lower---by a factor of $\sim 7$---than for linear regression with a standard deviation similarly lower by a factor of $\sim 3$.

\begin{table}[htbp]
    \centering
    \caption{Root mean squared error (RMSE) for ACA and linear regression, for the full model with all covariates.}
    \begin{tabular}{|l|l|l|}
    \hline
     \textbf{ACA}  & \textbf{Linear regression} \\ \hline
     4.36 $\pm$ 3.40      & 29.15 $\pm$ 10.02   \\ \hline
    \end{tabular}
    \label{tab:rmse_all}
\end{table}

However, as shown in Table \ref{tab:rmse_marginal} examining the marginal output that considers one feature at a time, linear regression outperforms ACA in RMSE by 2 to 7 mg/dl for breakfast, lunch, and dinner meals, while ACA slightly outperforms linear regression for analysis when all meals are pooled together.  The explanation: ACA, being a complex nonlinear regression, is more data-hungry than linear regression, and because it underperforms linear regression for a single meal but outperforms for three meals, it needs at most three times the data to have a lower RMSE than linear regression.

\begin{table}[htbp]
    \centering
    \caption{Root mean squared error (RMSE) for ACA and linear regression, for the marginal model considering one covariate at a time.}
    \begin{tabular}{|l|l|l|}
    \hline
    \textbf{Meal type} & \textbf{ACA}               & \textbf{Linear regression}     \\ \hline
    breakfast & 28.81 $\pm$ 16.2  & 26.27 $\pm$ 14.3      \\
    lunch     & 35.06 $\pm$ 18.0  & 32.62 $\pm$ 16.0      \\
    dinner    & 40.21 $\pm$ 26.1  & 33.60  $\pm$ 20.3      \\
    overall   & 37.21 $\pm$ 21.3  & 37.44 $\pm$ 21.4      \\ \hline
    \end{tabular}
    \label{tab:rmse_marginal}
\end{table}

The difference between ACA and the marginalized ACA --- that ACA itself produces very accurate representations of the data while the marginalization is substantially less accurate --- has important implications. First, this difference shows that there is substantial correlation between the covariates; this is not surprising because individual meals are combinations of food items, which in turn have combinations of macronutrients, suggesting that the macronutrients in a meal are not independent of each other. Second, it is clear that because of the systematic relationships between covariates, there is predictive information that we are not using to help people make decisions. The problem of course, is that the full portrait of how these covariates influence glycemic impact is a complex mathematical object. And to be useful in practice there is an imposed tradeoff that is not about algorithmic accuracy, but about human factors: we need the algorithm to be accurate but we must balance accuracy against the ability to use the output of the algorithm to make decisions. And this leads us to the third implication of the difference between the ACA and its marginalized form: we must find a way to exploit this yet-unused predictive information in a way that also allows for useful decision-making.

\subsubsection{Non-linear relationships}

\begin{figure}[htbp]
  \centering
  \includegraphics[width=4in]{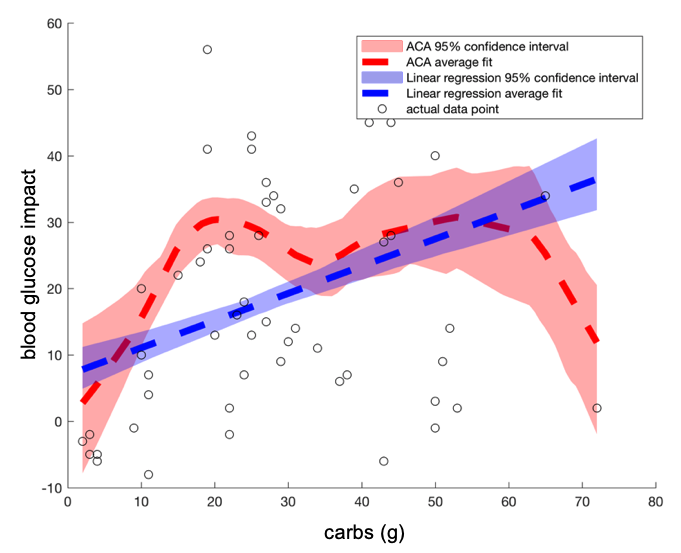}
  \caption{Comparison of ACA and linear regression for user A and the relationship between carbohydrates and BG impact, across all meals. In this case, ACA identifies a non-linear relationship, while linear regression does not.}
  \label{fig:nonlinearity}
\end{figure}

In some situations, ACA did identify non-linear relationships between macronutrients and BG impact, as shown in Figure \ref{fig:nonlinearity}. Because of the regularization built into ACA, most of the identified trends were linear, but some were non-linear. Non-linear relationships may be expected in some situations because of the complexity of BG dynamics. Linear regression, of course, would by definition never be able to find a non-linear relationship.

\subsubsection{Outliers and Errors}

\begin{figure}[htbp]
  \centering
  \includegraphics[width=4in]{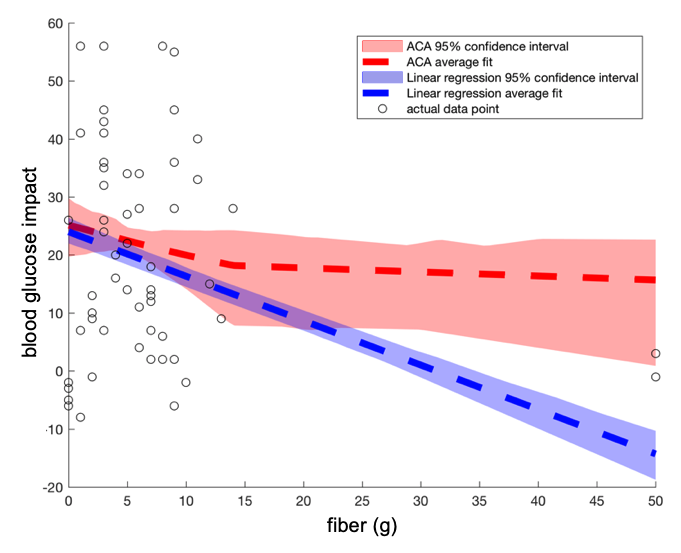}
  \caption{Comparison of ACA and linear regression for user A, and the relationship between fiber and BG impact, across all meals. ACA shows no trend leading out to the outlier data points with 50 grams of fiber, while linear regression continues a downward trend beyond what is reasonable.}
  \label{fig:outliers}
\end{figure}

When inspecting the plots, we found that some data sets had outliers that were clearly errors. For example, User A's data had two meals recorded with 50 grams of fiber. These data points are clearly errors not only because they are visibly separated from the rest of the data, but also because 50 grams was the default value for nutrient assessments by RDs, and 50 grams of fiber is an infeasible amount to eat in one sitting. The recommended amount of fiber is 38 grams per day for men, and 95\% of adults don't manage to eat the recommended amount of fiber; 50 grams of fiber would be over 3 cups of lentils. As shown in Figure \ref{fig:outliers}, linear regression is unable to ignore the outliers, and continues the downward trend beyond what is reasonable. ACA, on the other hand, also finds a slight downward trend in the non-outlier data, but evens out to be flat---showing no relationship---over the sparsely populated region before the outliers. The ACA is a more robust estimator \cite{huber2011robust} than linear regression.

\subsubsection{Uncertainty}

\begin{figure}[htbp]
    \centering
    \begin{subfigure}[t]{0.45\textwidth}
      \centering
      \includegraphics[width=2in]{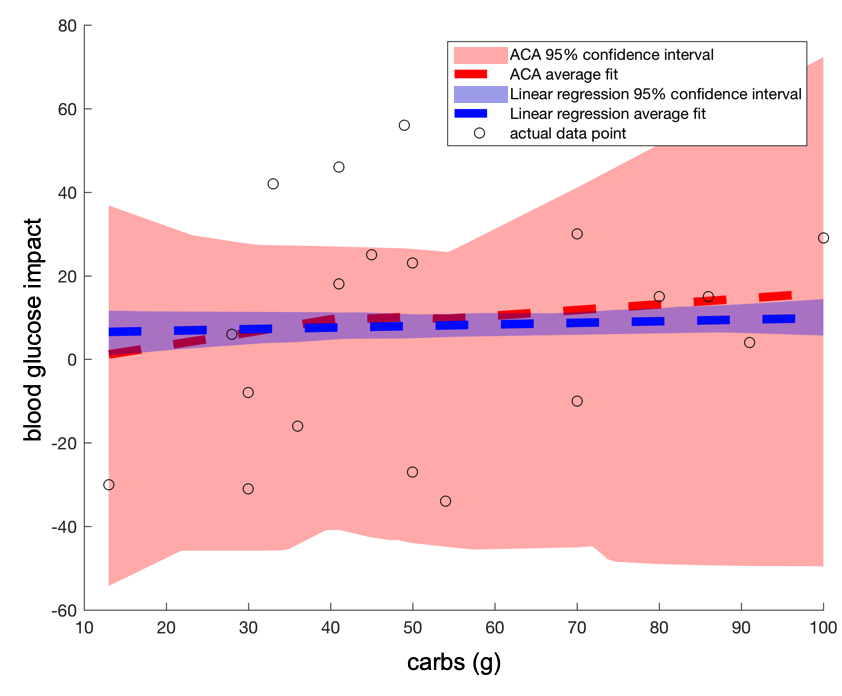}
    \end{subfigure}
    \begin{subfigure}[t]{0.45\textwidth}
      \centering
      \includegraphics[width=2in]{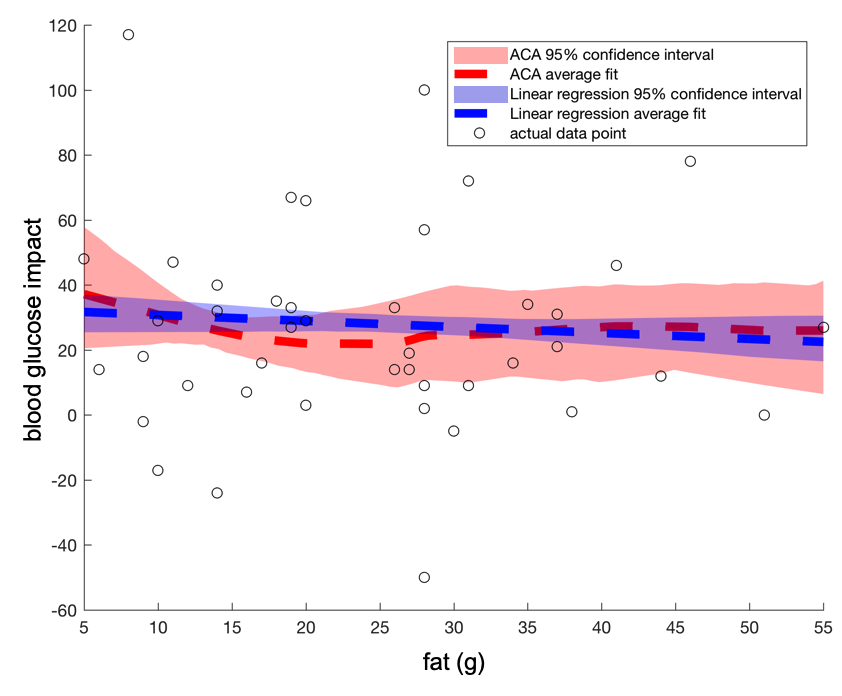}
    \end{subfigure}
    \caption{Comparison of ACA and linear regression for user B. On the left is the relationship between carbohydrates and BG impact for lunch meals. On the right is the relationship between fat and BG impact at dinner for the same user. On the left, ACA has wide confidence intervals, indicating uncertainty about the true relationship, while confidence intervals are narrower on the right. In contrast, linear regression has narrow confidence intervals in both figures.}
    \label{fig:uncertainty}
\end{figure}

One of the most drastic differences between ACA and linear regression was in the size and variability of the confidence intervals. Confidence intervals for ACA were broad, and varied in their width across data sets. In some instances, ACA would have a relatively narrow confidence interval, suggesting a higher degree of certainty in the identified trend. In other situations, though, ACA has broad confidence intervals, encapsulating most of the data sets, suggesting a low degree of confidence in the identified trend. On the other hand, the less flexible linear regression typically had narrow confidence intervals, regardless of the plausibility of the trend identified. See Figure \ref{fig:uncertainty} for a comparison of uncertainty between two data subsets for the same user.

In general, the confidence intervals were much wider and more expressive with ACA. As shown in Table \ref{tab:uncertainty}, more of the actual data points---by factors ranging from $2$ to $16$ with an average of $6$---fell within the confidence intervals for ACA than with linear regression.

\begin{table}[htbp]
    \centering
    \caption{Percent of data points within the 95\% confidence interval for attributable components analysis (ACA) and linear regression.}
    \label{tab:uncertainty}
    \begin{tabular}{|l|l|l|l|}
    \hline
                & \textbf{N}       & \textbf{ACA}               & \textbf{Linear Regression}            \\ \hline
    \textbf{User A}     &         &                   &            \\
    Breakfast   & 13      & 84.62\%           & 10.77\%    \\
    Lunch       & 10      & 28.00\%           & 2.00\%     \\
    Dinner      & 23      & 58.26\%           & 7.83\%     \\
    All meals   & 58      & 15.17\%           & 7.59\%     \\ \hline
    \textbf{User B}   &         &                   &            \\
    Breakfast   & 16      & 96.25\%           & 6.25\%     \\
    Lunch       & 19      & 52.63\%           & 8.42\%     \\
    Dinner      & 44      & 32.27\%           & 11.36\%    \\
    All meals   & 88      & 22.05\%           & 12.05\%    \\ \hline
   \textbf{All Users} (Mean $\pm$ SD)   &         &                   &            \\
    Breakfast   & 23 $\pm$ 16 & 62\% $\pm$ 21\%       & 11\% $\pm$ 8\% \\
    Lunch       & 21 $\pm$ 14 & 47\% $\pm$ 21\%       & 8\% $\pm$ 6\%  \\
    Dinner      & 24 $\pm$ 15 & 47\% $\pm$ 22\%       & 10\% $\pm$ 7\% \\
    All meals   & 82 $\pm$ 63 & 25\% $\pm$ 12\%       & 12\% $\pm$ 7\% \\ \hline
    \end{tabular}
\end{table}

\section{Discussion}

In this study, we explored the use of a method based on optimal transport theory to analyze patient-generated data. As compared to linear regression, we found that attributable components analysis (ACA) was able to identify non-linear relationships, was more robust to outliers, and offered more representative and accurate uncertainty estimates. These characteristics make ACA a good candidate to be used in the wild for decision support systems. For example, model output could be used in a tool to help clinicians deliver personalized coaching to patients with T2D, to automatically generate meal plans, or in a smartphone application that delivers personalized nutritional recommendations directly to patients.

Unlike post hoc data analysis, when datasets can be cleaned, curated, and processed, algorithms used in decision support systems need to run automatically without direct oversight using data with all their imperfections. Given the constraints of real self-monitoring data, the marginalized ACA preformed well.  But it is important to understand the modeling workflow we develop here, and its advantages and evaluation. We compared a simple regression, linear regression, to a complex nonlinear regression that was then simplified after the fact.  It seems that, given enough data, it is more productive to begin with a model capable of representing the structures in the data and have the features \textit{necessary for useful decision-making}, and then simplifying the model output as is required for practical decision support. Non-linear regressions are not always required or useful, and often a linear or logistic regression---as a sophisticated use of a simple tool---will be a better choice due to the needs of the application, e.g., \cite{levine2018methodological}.  Here we had substantial gains from basing the analysis in a more flexible tool, but also saw some drawbacks, all of which are noted below.

\textbf{Nonlinear relationships in data and decision support.} The ACA was able to identify non-linear relationships, which is important because of the complexity of BG dynamics and other systems in health. Importantly, ACA is also regularized to prevent overfitting, and the majority of relationships identified were linear. As discussed in \ref{regression2cds}, one approach to make regression output useful for decision support is to use a clinically meaningful threshold for BG impact to identify ranges of values to expect higher or lower BG impacts. Because ACA is non-linear, it can identify multiple ranges, but with linear regression, this approach would only identify 1 high and 1 low impact range. Distinct ranges may be more clinically meaningful.

\textbf{Robust estimation.}  ACA was more robust to outliers and erroneous data points than linear regression \cite{huber2011robust}. Data accuracy is a central concern in assessing the quality of electronic health data \citep{Hripcsak2013Next-generationRecords.,Weiskopf2013MethodsResearch}, especially for patient-generated health data, when patients are directly entering data points \citep{Codella2018DataData}. While rule-based or statistical methods can be used to detect and remove outliers, analytic approaches that are robust to outliers, like ACA, are still advantageous.

\textbf{Uncertainty quantification.} ACA offered broader and more representative and accurate uncertainty estimates than linear regression \cite{smith2013uncertainty}. It's important to represent and consider the confidence of the model for a given patient's data set. Uncertainty is intrinsic to the practice of medicine. If a model is going to be used for clinical decision support, representing the uncertainty can help clinicians appropriately weight the information against everything else they know about the patient \citep{Cabitza2019AMedicine,Cabitza2017UnintendedMedicine}. For patient-facing application, the certainty can help prioritize what is and is not shared with users.

\textbf{Reducing model flexibility for interpretability and decision-making.} Linear regression is rather interpretable, especially in one dimension.  A nonlinear regression like ACA, which models a distribution function that estimates glycemic response, is far less interpretable in its raw form; it often requires mathematical sophistication to interpret and is difficult to visualize due to the high-dimensional nature of the model.
While the full ACA model with all covariates outperformed linear regression, the quality of the fit dropped substantially when considering one covariate at a time in the marginal model given the data constraints. We focused on the marginal relationship between each covariate and glycemic impact because interpretability and actionability for decision support was a key objective: simultaneously making changes to multiple macronutrients is challenging for individuals to implement because of the cognitive burden and because behavior change is often grounded on incremental, achievable adjustments.

The poorer performance of the marginal model points to a tradeoff between accuracy and interpretability in machine learning tasks {\citep{Johansson2011Trade-offModeling}}. In this context, there is substantial information shared between covariates that is lost through marginalization.
{While the full model may be too complex for tractable interpretation, future work could explore marginalizing out fewer covariates, to examine the relationship between two covariates $H$ in relation to the quantity of interest $\bar{x}$, as opposed to a single covariate, as presented here. Three-dimensional surfaces can still be visualized and interpreted without adding unnecessary complexity, suggesting that this is a feasible direction for future work. In addition, such an approach could be employed alongside univariate marginalization when pre-hoc analysis suggests that two covariates share a great deal of information.}
{At the same time, there is a need for richer and more detailed model outputs in clinical characterization {\citep{Hripcsak2018}}, and future work could also explore ways to improve the interpretability of the full model with all covariates for use for decision support while still aligning with what clinician and patients need from a human factors standpoint.}

{Another important limitation is that ACA identifies patterns of association, but does not necessarily identify causal relationships in the data {\citep{blakely2019reflection}}. Future work could apply causal inference methodologies as an alternate path to increasing confidence that following a recommendation derived from the analysis will lead to improved BG impact for an individual.}

\textbf{Data limitations and machine learning.}  ACA, like all nonlinear regressions such deep learning or Gaussian process models, is more data-hungry than linear regression.  Meaning, the ACA requires more data \textit{to become} as accurate as linear regression.  Then, given enough data, the nonlinear regressions are generally more accurate or able to represent data than linear or other more rigid and simple regressions.  The flexibility of nonlinear regressions may not always be beneficial, depending on the application, but here the real question is of the limiting effect of data availability.  For example, while nonlinear regressions require more data than their more rigid counter-parts, not all methods have an equal hunger for data.  And here, because of the nature of our experiment, we have a window into how hungry ACA is compared to linear regression: ACA underperforms linear regression for a single meal but outperforms for three meals for most patients, meaning it needs at most three times the data to have a lower RMSE than linear regression. This is important because the whole point of personalized analysis of the BG impact of nutrition is to use an individual's data to estimate a model and provide decision support.  Moreover, because health states change, models must be re-estimated periodically---potentially every 3-6 weeks---and so to be impactful, the model must perform with self-monitoring data collected on the order of weeks. Given its lower RMSE than ACA in the marginal case, linear regression could still be useful for decision support, as its results are similarly interpretable. However, for the reasons discussed above, augmentations would be necessary. For example, to improve robustness to outliers or apply statistical approaches like anomaly detection to remove possibly erroneous data points. In addition, linear regression would benefit from improved uncertainty estimates, or it would be difficult to determine when signals are clinically meaningful or actionable. For these reasons, devising methods to boost the impact of finite yet personal self-management data will be crucial.

\textbf{Human-centered data collection limitations.} The data available for analysis in realistic settings represents a limitation. T2D self-monitoring data is effortful for individuals to collect, and data sets are often small. As discussed in the feature selection section, BG readings before and two hours after each meal don't fully capture fluctuations in BG. Continuous glucose monitors (CGM) could provide more granular and accurate data for machine learning, but are not standard care for T2D, making them prohibitively expensive for most patients. Still, prior research has demonstrated the feasibility of similar data sets to make accurate prediction of BG values \citep{Albers2017PersonalizedAssimilation}. In the future, self-monitoring and other data sources from more individuals could be combined to find patterns of individuals with similar characteristics who share similar BG dynamics, for example by utilizing microbiome or electronic health record data \citep{Zeevi2015}. Notably, while researchers have been successful in predicting blood glucose and making nutrition recommendations to improve BG control, their models relied on extensive, and complete data about each individual \citep{Zeevi2015}. Personalized nutrition recommendations from self-monitoring data would be considerably more scalable for a large population.

In conclusion, this work presents initial progress in applying machinery from optimal transport theory to address important problems in machine learning with patient-generated health data.

\section{Data Sharing Statement}
The data used in this project is considered protected health information (PHI) and therefore cannot be made openly available for general use. The authors are open to collaborating with researchers, to build on or reproduce the methods and analyses described in this work. Please email the corresponding author if you are interested in collaborating or reproducing this work (Elliot G Mitchell: egm2143@columbia.edu)

\section{References}

\bibliography{references}

\appendix

\section{Attributable Components Analysis}
\label{appendix_a}

\subsection{Modes of variability}

Given a set of $m$ observations $\{(x^i, z^i)\}_{i=1}^m$ of the variable of interest $x$ and the $L$ covariates $z_l$, we seek to estimate the conditional mean $\bar{x}(z)$. We will assume throughout that $x$ assumes real values; this is a reasonable assumption given that $x$ represents glycemic impact that we define here as the difference between two real numbers, pre- and postprandial blood glucose measurements. If we instead specified $x$ as a vector, the $j$'th component of their mean is the mean of the $j$'th component, so there is no loss of generality in considering one dimension of $x$ at a time. In this application, $x$ only has one dimension. We will leave the specification of the allowable variable types for each covariate $z_l$ temporarily open.

The conditional mean can be characterized as the minimizer of the variance:
\begin{equation}
  \bar{x}(z) = \arg \min_{f} \sum_i \left\|x^i - f(z^i) \right\|^2
  \label{xm}
\end{equation}
over a proposed family of functions $f(z)$. We would like our specification of this family of functions to satisfy some properties:

\begin{enumerate}
  \item The family should be big enough to accurately represent complex dependencies of $\bar{x}$ on $z=(z_1, \ldots, z_L)$, while at the same time constrained so as not to overfit the data.

    \item The procedure should be applicable to covariates $z_l$ of quite arbitrary type.

  \item It should be \emph{interpretable}, in the sense that one should be able to compute with ease the \emph{marginal} dependence on $f$ on some subset of the $z_l$, averaging over the others.

  \item Performing the minimization in (\ref{xm}) should be computationally effective.

\end{enumerate}

The choice made in \cite{tabak-trigila2018AC} is to approximate the multivariable function $f(z)$ by the superposition of $d$ products of functions $V(l)^k$ of the individual covariates $z_l$:

\begin{equation}
 f(z) \approx \sum_{k=1}^d \prod_{l=1}^L V(l)^k\left(z_l\right).
 \label{Modes}
\end{equation}
This can be thought as an extension of the low-rank factorization of matrices
$$ A_i^j \approx \sum_{k=1}^d u^k_i v^k_j $$
from matrix entries considered as functions of the row and column, to tensors of arbitrary order and variables of arbitrary type.
Two explicit examples with real covariates but functions $V(l)^k$ pre-assigned except for a global multiplicative factor are the power series
$$ f(z) \approx \sum_{k} a_k \prod_{l=1}^L {z_l}^{s^k_l}, \quad z \in \R^L, \quad s^k_l \in \mathbb{N}$$
and the Fourier series
$$ f(z) \approx \sum_{k}  a_k e^{i \sum_l  \xi^k_l z_l}, \quad z_l \ \hbox{$2\pi$-periodic,} \quad \xi^k_l \in \mathbb{Z}.$$

In the context of low rank factorization, particularly when it is re-arranged so that both the $\left\{u^k\right\}$ and $\left\{v^k\right\}$ are orthogonal sets of vectors (i.e. principal component analysis),  each $k$ represents a \emph{component} of variability,  typically sorted by the fraction of total variability that each component explains. In Fourier analysis, a linear case, one speaks of Fourier \emph{modes}. Because we are not anchored to a particular functional form for the $f$'s we will refer to each product $\prod_{l=1}^L V(l)^k\left(z_l\right)$ as a \emph{mode of variability} of $\bar{x}(z)$, and think of it as a pattern of dependence on $z$ to be extracted from the data that explains a significant fraction of the variability of $x$.

Under the proposal in (\ref{Modes}), the conditional expectation problem in (\ref{xm}) reduces to
\begin{equation}
\min_{V} L = \sum_i \left(x^i - \sum_{k=1}^d \prod_{l=1}^L V(l)^k(z_l^i) \right)^2
 \label{min_var}
\end{equation}
over the degrees of freedom available in the specification of the functions $V(l)^k(z)$. We discuss next how to specify these functions.

\subsection{Hard and soft assignments (coping with missing values), grids and prototypes}

If the $z_l$ are categorical variables, such as the rows and columns in low-rank matrix factorization, we can assign an integer $j \in [1,\ m(l)]$ to each of the values they can adopt. Then each $V(l)$ is fully described by a matrix with components $V(l)^k_{j} = V(l)^k(j)$, and we can write
\begin{equation}
   V(l)^k(z_l^i) = \sum_j \alpha(l)_i^j V(l)^k_j,
   \label{alphas}
\end{equation}
where $\alpha(l)_i^j = 1$ when $z_l^i = j$ and $\alpha(l)_i^j = 0$ otherwise.

Since $L$ is quadratic in each $V(l)$, one can perform the minimization of (\ref{xm})  through an alternating direction methodology, minimizing $L$ alternatively over each $V(l)$, which yields the updating rule
\begin{equation}
  V(l)_j = \left(\sum_{i \in I_j} x^i \prod_{b\ne l} V(b)_{z_b^i} \right) \left[\sum_{i \in I_j} \left(\prod_{b\ne l} V(b)_{z_b^i} \right)^T \left(\prod_{b\ne l} V(b)_{z_b^i} \right) \right]^{-1},
\end{equation}
where
$$ I_j = \left\{i : z_l^i = j\right\}. $$

We can extend the applicability of (\ref{alphas}) to situations where the value of $z_l$ in some or all observations are not known with certainty. Then $\alpha(l)_i^j$ is no longer a binary variable with values zero or one, but represents instead the probability that $z_l^i$ adopts the value $j$. This \emph{soft assignment} satisfies
\begin{equation}
  \forall i, \ \alpha(l)_i^j \ge 0, \quad \sum_j \alpha(l)_i^j = 1.
  \label{convexity}
\end{equation}
This allows us a means of naturally accommodating both measurement uncertainty and one pathway for coping with missing data within a covariate, $z_l$.

In the event that the covariates, $z_l$, are real, we can extend (\ref{alphas}, \ref{convexity}) by adopting a grid; here we adopt a grid ${z_g}(l)^j$ and define $V(l)_j^k = V(l)^k\left({z_g}(l)^j\right)$. Then, performing a piecewise linear interpolation, one can assign to each observation $z_l^i$ values of $\alpha(l)_i$ such that
$$
 z_l^i = \sum_j \alpha(l)_i^j {z_g}(l)^j $$
and write again
$$ {V}(l)^k\left({z_l^i}\right) = \sum_j \alpha(l)_i^j V(l)_j^k. $$
Here the $\alpha$ satisfy, in addition to (\ref{convexity}), the condition that at most two $\alpha(l)_i^j$ differ from zero for each value of $i$, the ones corresponding to the two grid points surrounding $z_l^i$.

Finally, the formulation in (\ref{alphas}, \ref{convexity}) can be further extended to any type of covariate $z_l$ that admits a norm, via prototypal analysis \citep{prototypes}. In this case the grid
${z_g}(l)^j$ is replaced by the \emph{prototypes} $y_l^j$, which are optimal convex combinations of the $z_l^i$,
$$ y_l^j = \sum_i \beta(l)_i^j z_l^i, \quad \beta(l)_i^j \ge 0, \quad \sum_i \beta(l)_i^j =1,$$
where $\alpha(l)$ and $\beta(l)$ solve the following minimization problem:

\begin{equation}
 \alpha(l),\beta(l) = \argmin \sum_i \left\|z_l^i - \sum_j \alpha(l)_i^j  y_l^j \right\|^2  + \lambda \sum_{i,j} \|z_l^i-y_l^j\|^2 \alpha(l)_i^j,
\label{prototypes}
\end{equation}
$$ y_l^j = \sum_i \beta(l)_i^j z_l^i, \quad \alpha(l)_i^j \ge 0, \quad \beta(l)_i^j \ge 0, \quad \sum_j \alpha(l)_i^j  = \sum_i \beta(l)_i^j =1.$$

The interpretation is the following: in archetypal analysis \citep{archetype_analysis}, we seek points $y_l^j$ within the convex hull of the $z_l^i$ such that the latter can be well approximated by convex combinations of the former. What the prototypes add is the penalization term with strength $\lambda$, which favors expression for the $z_l^i$ that are \emph{local}, i.e. involve only nearby $y_l^j$, as with the piecewise linear expansions adapted to a grid.

\subsection{Smoothness and bounded variability}

As the grids ${z_g}(l)^j$ become finer or the number of prototypes $y_l^j$ grows to permit a more accurate representation of the $V(l)(z_l)$, the risk of overfitting the data also increases. To avoid this, one can enforce smoothness on $V(l)(z_l)$, for instance by penalizing the squared norm of a finite difference approximation to its gradient. A general form for such a penalization term $P$ is
$$ P = {V(l)^k}' C^l V(l)^k, $$
where the matrix $C^l$ encodes the specific penalization used, such as the squared norms of first or second derivatives. A similar term can be used for categorical variables, encoding into $C^l$ their variance, to bound the amount of variability that they can explain. Then the full problem adopts the form
\begin{eqnarray}
\min_{V} \sum_i \left(x^i - \sum_k \prod_{l \in L} \sum_j \alpha(l)_i^j V(l)_j^k \right)^2 + \nonumber\\
\sum_{l=1}^L \lambda_l  \sum_k \left(\prod_{b \in L, b\ne l} \|V(b)^k\|^2\right)
{V(l)^k}' C^l V(l)^k.
\label{SupComp4}
\end{eqnarray}

The inclusion of the  products of squares of the norms of the $V(b)^k$ as pre-factors to the penalty terms follows from the need to make the objective function invariant under re-scalings of the $V(l)^k$ that preserve their product. Without these, the penalty terms could be made arbitrarily small by rescaling each $V(l)$ while preserving their product, assigning large amplitudes to those $V(l)$ that can explain little or no variability, and can therefore be taken as constants so that the corresponding ${V(l)^k}' C^l V(l)^k$ vanishes.

Notice that the objective function in (\ref{SupComp4}) is still quadratic in each $V(l)$, and so can be solved through an alternative direction methodology that finds the optimal matrix $V(l)$ explicitly given the current values of the $V(b)$ for $b\ne l$.

\end{document}